# Measuring x-ray polarization in the presence of systematic effects: Known background

Ronald F. Elsner[*], Stephen L. O'Dell, and Martin C. Weisskopf

NASA Marshall Space Flight Center, Astrophysics Office, Huntsville, AL 35812

## ABSTRACT

The prospects for accomplishing x-ray polarization measurements of astronomical sources have grown in recent years, after a hiatus of more than 37 years. Unfortunately, accompanying this long hiatus has been some confusion over the statistical uncertainties associated with x-ray polarization measurements of these sources. We have initiated a program to perform the detailed calculations that will offer insights into the uncertainties associated with x-ray polarization measurements. Here we describe a mathematical formalism for determining the 1- and 2-parameter errors in the magnitude and position angle of x-ray (linear) polarization in the presence of a (polarized or unpolarized) background. We further review relevant statistics—including clearly distinguishing between the Minimum Detectable Polarization (MDP) and the accuracy of a polarization measurement.

## 1. INTRODUCTION

Only a few experiments have conducted x-ray polarimetry of cosmic sources. About 40 years ago, several rocket observations[1] measured x-ray polarization from the Crab Nebula. The x-ray polarimeter aboard the Orbiting Solar Observatory (OSO-8) confirmed this result with a 19-sigma detection[2] (P = 19.2% ± 1.0%), thus proving the synchrotron origin of the x-ray emission. Data from OSO-8 also provided upper limits to polarization for a few x-ray-bright targets. The key challenge for broad-band x-ray polarimetry has always been to make an instrument that is highly responsive to polarized radiation but does not suffer from background or other systematic effects. In the past decade, development of a class of polarimeters based upon the photoelectric effect[3,4,5] and placed at the focus of an x-ray telescope has produced a renewed interest in x-ray polarimetry of cosmic sources.

A realistic assessment of understanding prospects for performing meaningful measurements of the linear polarization of astronomical x-ray sources critically depends upon understanding uncertainties produced by statistical fluctuations in polarization data, especially important in the presence of systematic effects. Unfortunately there has been some confusion regarding figures of merit used to compare the relative sensitivity of different instruments, and the impact of systematic effects that cannot simply be "calibrated out" without potentially adverse impact on the instrument's capability for obtaining scientifically meaningful measurements.

We previously addressed[6] statistical uncertainties associated with polarization measurements of astronomical sources. Subsequently, we undertook a detailed program to investigate the impact of systematic effects on polarization measurements. Here (§2) we here summarize calculations of the statistical uncertainty in measuring linear polarization in the presence of a (polarized or unpolarized) background. Next (§3) we illustrate results for a hypothetical measurement of the polarization of the Crab Pulsar as a function of pulse phase, for polarimeters that may or may not spatially resolve the pulsar from the nebula, which serves as a polarized background. We conclude (§4) with a summary of current results and a brief discussion of applying our formalism to other systematic effects that affect the polarimeter's sensitivity and accuracy.

---

[*] Contact author: Ronald.F.Elsner@nasa.gov; phone +1 256-961-7765; fax +1 256-961-7522
Postal address: NASA/MSFC/ZP12, 320 Sparkman Drive NW, Huntsville, AL 35805-1912 USA

## 2. STATISTICS OF POLARIZATION MEASUREMENTS

A polarimeter measures linear polarization by determining the amplitude and phase of modulation with azimuthal angle $\varphi$ by analyzing the data for a $2\varphi$ signature. As the calculated polarization $\varpi$ depends upon the directly-measured (fractional) modulation amplitude $a$ and an instrument-dependent modulation factor $\mu$ (the modulation amplitude for a 100%-polarized beam), we shall work with the modulation amplitude $a$ throughout this paper. After finding $a$, we can then calculate the fractional polarization using

$$\varpi = a/\mu. \qquad (1)$$

The probability density function (PDF) specifies the distribution of measured parameters about their true values. For example, the bivariate PDF for a measurement of the modulation amplitude $a$ and position angle $\psi$, given their expectation values (denoted by a caret), is symbolically

$$p_{a,\psi}(a,\psi \mid \hat{a},\hat{\psi}). \qquad (2)$$

In Appendix A, we sketch the calculation of the PDF and derived statistical properties—minimum detectable amplitude (MDA) and univariate uncertainties ($\delta a$, $\delta\psi$) in parameters ($a$, $\psi$)—for a polarized source in the presence of a known (polarized or unpolarized) background. Here we summarize the primary results. Along the way, we offer a few comments to clarify the results and their limitations.

The starting point for deriving the bivariate PDF for the polarization of a source in the presence of a known (polarized or unpolarized) background, is the azimuthal density of $\Delta\chi^2$, the difference in $\chi^2$ between two sets of model parameters describing the (polarization signature) $2\varphi$ modulation of counts with azimuth $\varphi$.

$$\frac{d}{d\varphi}\left[\Delta\chi^2(\varphi)\right] = \{C_S[a_S\cos(2(\varphi-\psi_S)) - \hat{a}_S\cos(2(\varphi-\hat{\psi}_S))] + C_B[a_B\cos(2(\varphi-\psi_B)) - \hat{a}_B\cos(2(\varphi-\hat{\psi}_B))]\}^2 \\ \div \{2\pi(C_S[1+\hat{a}_S\cos(2(\varphi-\hat{\psi}_S))] + C_B[1+\hat{a}_B\cos(2(\varphi-\hat{\psi}_B))])\}. \qquad (3)$$

Here, the polarization parameters are the (fractional) modulation amplitude ($a_S$ and $a_B$) and position angle ($\psi_S$ and $\psi_B$) of the total counts from the source ($C_S$) and from the background ($C_B$), respectively. The caret over each corresponding symbol denotes the "true" expectation value of that parameter.

First, in order to proceed mathematically and to obtain the standard expression for the bivariate PDF for the source polarization parameters, we assume that the statistical variance is azimuthally uniform—i.e., that the denominator of Equation (3) does not vary with $\varphi$. While this is strictly valid only for the double-null hypothesis that the source and background are each unpolarized, it is a reasonable approximation provided that the modulation amplitudes are small ($a^2 \ll 1$). Fortunately for the mathematics but unfortunately for the measurements, this is usually the case! With this assumption, it is straightforward to integrate over azimuth $\varphi$ to obtain $\Delta\chi^2$.

Second, for simplicity, we consider the background to be perfectly known—$a_B \to \hat{a}_B$ and $\psi_B \to \hat{\psi}_B$—such that the remaining parameters of interest are $a_S$ and $\psi_S$. This reduces the quadrivariate PDF to a bivariate PDF

$$p_{a_S,\psi_S}(a_S,\psi_S \mid \hat{a}_S,\hat{\psi}_S) = \frac{a_S}{\pi\sigma_{a_S}^2}\exp\left[-\frac{1}{2}\Delta\chi^2(a_S,\psi_S \mid \hat{a}_S,\hat{\psi}_S)\right] \\ \approx \frac{a_S}{\pi\sigma_{a_S}^2}\exp\left[-\frac{(a_S-\hat{a}_S)^2 + a_S\hat{a}_S[2\sin(\psi_S-\hat{\psi}_S)]^2}{2\sigma_{a_S}^2}\right], \qquad (4)$$

which is normalized over the intervals $0 \le a_S < \infty$ and $0 \le \psi_S < \pi$.

Third, note that the modulation amplitude and the position angle of the background polarization no longer appear in the expressions for $\Delta\chi^2$ and for the consequent PDF, in that the background is taken to be perfectly known. However, background counts still contribute to the statistical noise in measuring the modulation amplitude. Specifically, the standard deviation for determining the modulation amplitude

$$\sigma_{a_S} = \sqrt{2}\,\tilde{\sigma}_{a_S} \equiv \sqrt{2(C_S + C_B)}/C_S , \qquad (5)$$

with $\tilde{\sigma}_{a_S} \equiv \sqrt{(C_S + C_B)}/C_S$, the standard deviation for fitting the modulated data. This factor of $\sqrt{2}$, which has caused some confusion, essentially results from the ratio of the amplitude of modulation to its RMS (root mean square) value.

The familiar (Rice distribution) univariate PDF for the modulation amplitude $a_S$ follows Equation (4) by integrating over position angle $\psi_S$.

$$p_{a_S}(a_S \mid \hat{a}_S) \approx \frac{a_S}{\sigma_{a_S}^2} \exp\left[-\frac{\left(a_S^2 + \hat{a}_S^2\right)}{2\sigma_{a_S}^2}\right] I_0\left[\frac{a_S \hat{a}_S}{\sigma_{a_S}^2}\right], \qquad (6)$$

where $I_0(x)$ is the modified Bessel function of the first kind and order 0.

The statistical metrics for polarization all depend upon $\sigma_{a_S}$ (Equation 5), the standard deviation for the uncertainty in measuring the modulation amplitude $a_S$. Integration of $p_{a_S}(a_S \mid 0)$ —Rayleigh distribution—yields the well-known formula for the minimum detectible amplitude $\text{MDA}_{\text{CL}}$ at confidence level CL.

$$\text{MDA}_{\text{CL}} \equiv \sqrt{-2\ln(1-\text{CL})}\,\sigma_{a_S} = \sqrt{-2\ln(1-\text{CL})}\,\sqrt{2(C_S + C_B)}/C_S . \qquad (7)$$

The S-sigma univariate uncertainties in the modulation amplitude and in the position angle, respectively, are

$$\delta a_S(\text{S}) = \text{S} \times \sigma_{a_S} = \text{S} \times \sqrt{2(C_S + C_B)}/C_S ; \qquad (8)$$

$$\delta\psi_S(\text{S}) = \text{S} \times \frac{\sigma_{a_S}}{2a_S} = \text{S} \times \frac{\sqrt{2(C_S + C_B)}/C_S}{2a_S} . \qquad (9)$$

It is important to recognize that the criterion for an accurate measurement is much more demanding than that for a secure detection. For example, if polarization is detected with a modulation amplitude equal to $\text{MDA}_{99}$, the measurement is only at about the 3-sigma level ($\sqrt{-2\ln(1-\text{CL})} \to 3.03$ for CL=0.99). Achieving a measurement accuracy of 10% (S=10), say, would require *an order of magnitude* more counts and correspondingly longer observation time. If one also seeks to measure accurately the spectral or temporal dependence of the polarization, the demands on observing time are even more severe.

## 3. EXAMPLE: CRAB PULSAR

We now apply the results to measurement of the polarization of the Crab Pulsar in the presence of the polarized signal from the Crab Nebula, which acts as a polarized background to measurement of the Pulsar. For this calculation, we assume the following:

1. The polarimeter is an imaging system with negligible instrumental background for the measurement.
2. Most of the Pulsar emission is contained in a resolution element $\Omega_P$, with total number of Pulsar counts $C_P$ in the resolution element and in the energy band of interest.

3. The number of counts from the Nebula in the resolution element is roughly $(\Omega_P / \Omega_N) C_N$, where $C_N$ is the total number of counts from the Nebula in the solid angle $\Omega_N$ ($\approx 1$ arcmin$^2$) and in the energy band of interest.
4. No additional systematic effects are significant.

Under these assumptions, the minimum detectable polarization at a confidence level CL is

$$\mathrm{MDP}_{\mathrm{CL},P} = \frac{\mathrm{MDA}_{\mathrm{CL},P}}{\langle\mu\rangle_P} = \frac{2\sqrt{-\ln(1-\mathrm{CL})}}{\langle\mu\rangle_P} \frac{\sqrt{C_P + (\Omega_P/\Omega_N)C_N}}{C_P}, \qquad (10)$$

such that the 99%-confidence (CL = 0.99) minimum detectable polarization is

$$\mathrm{MDP}_{99,P} = \frac{\mathrm{MDA}_{99,P}}{\langle\mu\rangle_P} = \frac{4.292}{\langle\mu\rangle_P} \frac{\sqrt{C_P + (\Omega_P/\Omega_N)C_N}}{C_P}. \qquad (11)$$

In general, the expected instantaneous count rate $\dot{C}(t)$ of photons in the energy range $E_l - E_u$ depends upon the energy-dependent net effective area $A_\mathrm{eff}(E)$ of the telescope system and the photon spectral flux $N_E(E,t)$ from the source:

$$\dot{C}(t) = \int_{E_l}^{E_u} dE \, \{A_\mathrm{eff}(E)[N_E(E,t)]_P\}. \qquad (12)$$

In order to perform phase-resolved polarimetry of a pulsar, we must integrate count rates over a pulse-phase range $\phi_l$–$\phi_u$, folded on the period $t_P$. The number of pulsar counts in the energy range $E_l - E_u$, over the phase range $\phi_l$–$\phi_u$, for an observation of total duration $t_\mathrm{obs}$ is then

$$C_P = \frac{t_\mathrm{obs}}{t_P} \int_{t_P\phi_l}^{t_P\phi_u} dt \, \{\dot{C}_P(t)\} = \frac{t_\mathrm{obs}}{t_P} \int_{t_P\phi_l}^{t_P\phi_u} dt \int_{E_l}^{E_u} dE \, \{A_\mathrm{eff}(E)[N_E(E)]_P\}. \qquad (13)$$

As the net duration of the observation over the selected pulse-phase range is $\Delta t = t_\mathrm{obs}\Delta\phi = t_\mathrm{obs}(\phi_u - \phi_l)$, the corresponding number of (temporally constant) nebular counts accumulated over that pulse-phase range is

$$C_N = \frac{t_\mathrm{obs}}{t_P} \int_{t_P\phi_l}^{t_P\phi_u} dt \int_{E_l}^{E_u} dE \, \{A_\mathrm{eff}(E)[N_E(E)]_N\} = t_\mathrm{obs}(\phi_u - \phi_l) \int_{E_l}^{E_u} dE \, \{A_\mathrm{eff}(E)[N_E(E)]_N\}. \qquad (14)$$

Finally, conversion of the (directly measurable) fractional modulation amplitude $a_P$ into a fractional polarization $\varpi_P$ (or an MDA into an MDP) requires knowledge of the polarimeter's modulation factor $\mu(E)$, appropriately averaged over photon energy $E$ for the pulsar's spectrum over the selected pulse-phase range:

$$\langle\mu\rangle_P = \frac{t_\mathrm{obs}}{t_P} \int_{t_P\phi_l}^{t_P\phi_u} dt \int_{E_l}^{E_u} dE \, \{\mu(E) A_\mathrm{eff}(E)[N_E(E,t)]_P\} / C_P = \frac{\int_{t_P\phi_l}^{t_P\phi_u} dt \int_{E_l}^{E_u} dE \, \{\mu(E) A_\mathrm{eff}(E)[N_E(E,t)]_P\}}{\int_{t_P\phi_l}^{t_P\phi_u} dt \int_{E_l}^{E_u} dE \, \{A_\mathrm{eff}(E)[N_E(E,t)]_P\}}. \qquad (15)$$

As an example, we adopt the effective area $A_\mathrm{eff}(E)$ and modulation factor $\mu(E)$ as functions of photon energy $E$ for an imaging x-ray polarimeter explorer (IXPE) proposed in 2007-2008. The response of the polarimeter (Figure 1) governs the effective energy range ($E_l$–$E_u$) of the measurement, which we set to 1.6–6.0 keV.

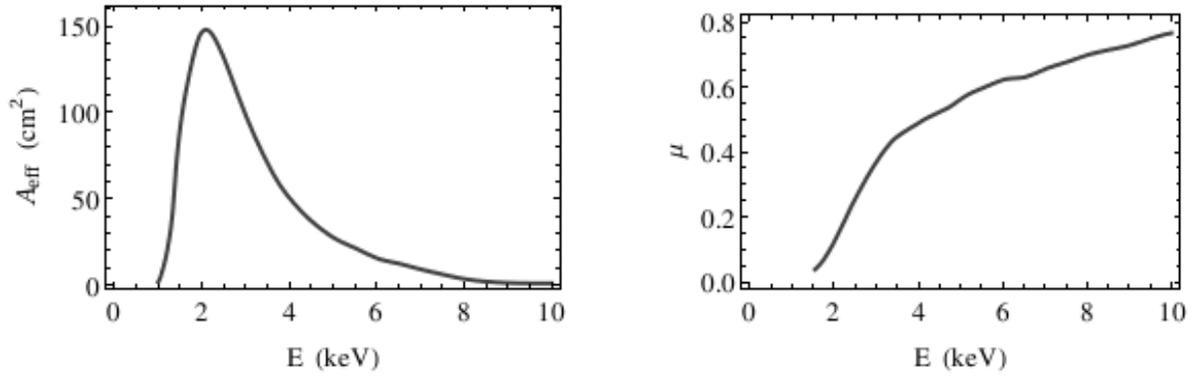

Figure 1: Performance characteristics of the IXPE polarimeter, as functions of the photon energy $E$. The left panel displays its effective area $A_{\text{eff}}(E)$; the right panel, its polarization modulation factor $\mu(E)$—i.e., the fractional modulation amplitude for a 100% linearly polarized beam.

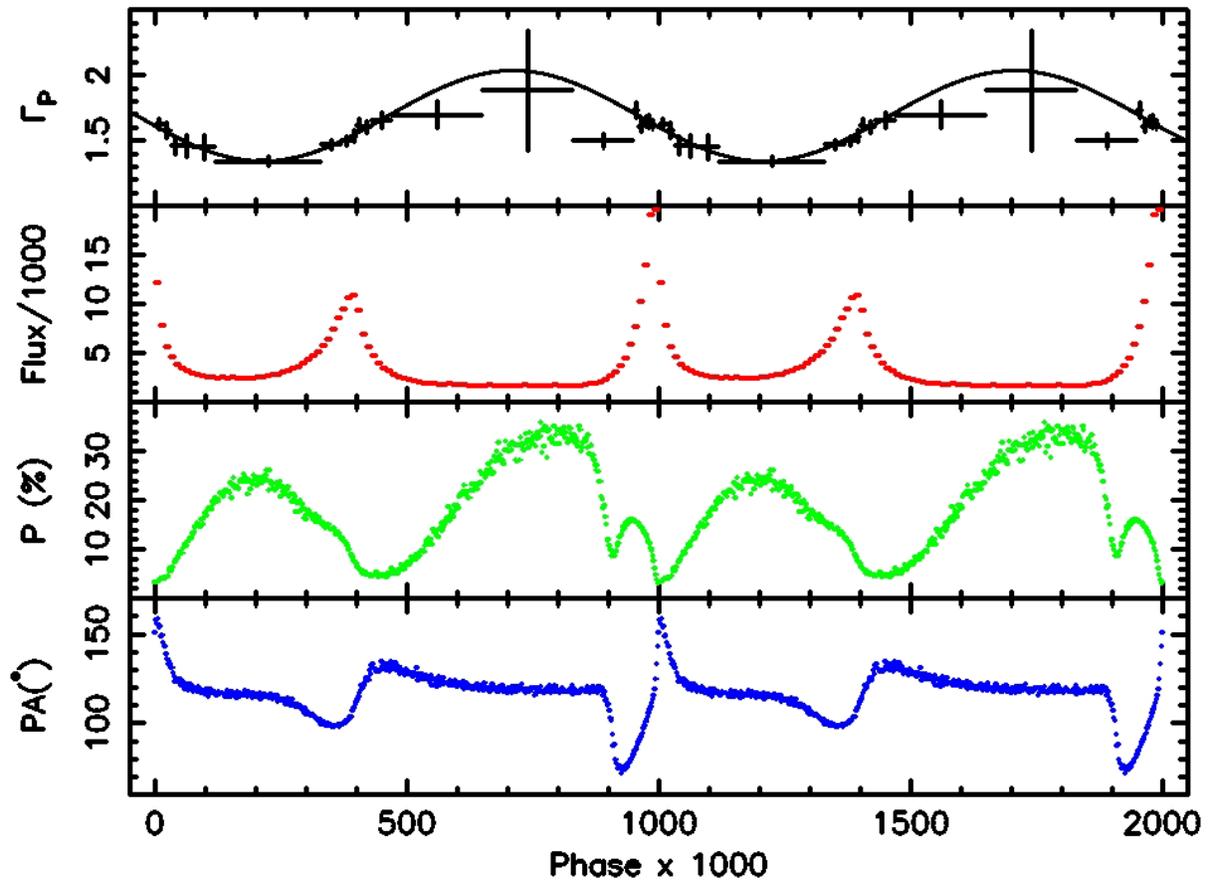

Figure 2: Phase-resolved multi-band data for the Crab Pulsar. The top two plots[8] display the x-ray photon index $\Gamma_x$ and photon spectral flux $N_E$ at $E_x = 1$ keV; the bottom two plots[7], the visible-light polarization degree and position angle. The plots show data folded on the pulsar's period ($t_P = 34$ ms) and repeated over two cycles for clarity. (Reproduced from reference [8].)

For the phase-resolved x-ray spectra of the Crab Pulsar (Figure 2), we adopt the values for hydrogen column density $N_H$, power-law norm $N_E$ (photon spectral flux at $E_x = 1$ keV), and photon index $\Gamma_x$ from *Chandra* data[8]. For the Crab Nebula, we use the same hydrogen column density as for the Pulsar, but adopt the x-ray power-law norm and photon index from XMM-*Newton* data[9].

Figure 3 illustrates the dependence of $MDP_{99,P}$ upon total observation time $t_{obs}$, for three features in the Crab Pulsar's pulse profile (Figure 2): inter pulse (phase range 0.03–0.33); secondary pulse (phase range 0.33–0.43); and primary pulse (phase range 0.95–0.03). The plot compares the $MDP_{99,P}$ for these three features, using different assumptions about the imaging performance of the polarimeter: non-imaging (Nebula unresolved); 15-arcsec imaging; and perfect imaging. Clearly, an imaging capability greatly enhances the sensitivity for detecting polarization of point source against a background (either nebular or instrumental).

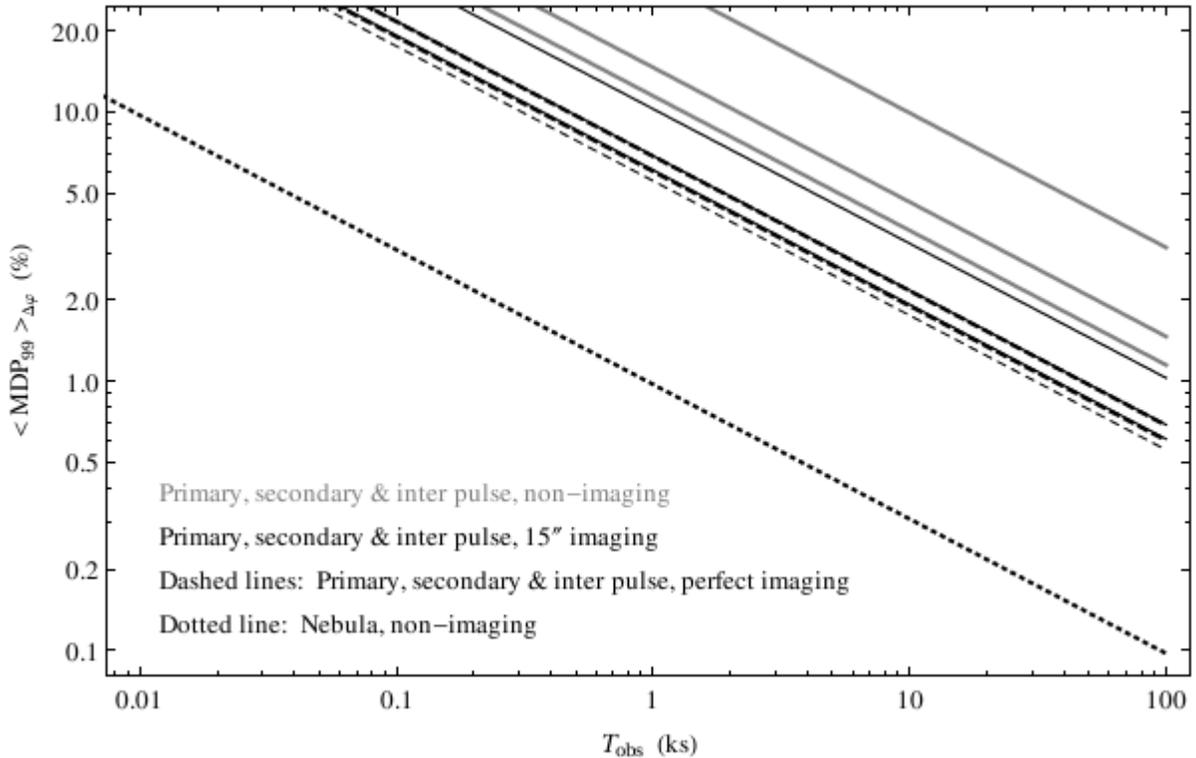

Figure 3: 99%-confidence Minimum Detectible Polarization ($MDP_{99}$) for specified pulse-phase intervals of the Crab Pulsar, showing importance of imaging, especially away from pulse peaks. In order of decreasing MDP (increasing sensitivity), each family of curves shows the x-ray $MDP_{99,P}(t_{obs})$ for the inter pulse (0.03–0.33), secondary pulse (0.33–0.43), and primary pulse (0.95–0.03). The solid-grey, solid-black, and dashed-black lines denote non-imaging, 15-arcsec-imaging, and perfect-imaging polarimeters, respectively.

Similarly, Figure 4 exhibits the dependence of the uncertainty $\sigma_{\varpi_P}$ in a polarization measurement $\varpi_P$ upon total observation time $t_{obs}$, for $\Delta\phi = 0.01$ phase bins near three features in the Crab Pulsar's pulse profile (Figure 2): mid inter pulse ($\phi_{IP} = 0.15$); secondary-pulse peak ($\phi_{SP} = 0.40$); and primary-pulse peak ($\phi_{PP} = 0.00$). The plot compares the $\sigma_{\varpi_P}$ for these three features, for different assumptions about the imaging performance of the polarimeter: non-imaging (Nebula unresolved); 15-arcsec imaging; and perfect imaging. Clearly, an imaging capability greatly enhances the sensitivity for detecting polarization of point source against a background (either nebular or instrumental). For a pulsar, rapid changes in polarization amplitude and position angle (Figure 2 bottom two plots) necessitate accumulation of data in rather narrow pulse-phase bins, in order to obtain an accurate measurement of $\varpi_P$ and $\psi_P$ and their pulse-phase dependence. Note that for a polarization measurement of 10% accuracy, the polarization amplitude would need to be 10-sigma. For example, to measure 5% polarization with 10% accuracy requires a measurement uncertainty $\sigma_{\varpi_P} < 0.5\%$. Even for an x-ray source as bright as the Crab pulsar, accurate phase-resolved polarimetry will be challenging. Without imaging, it would be barely feasible.

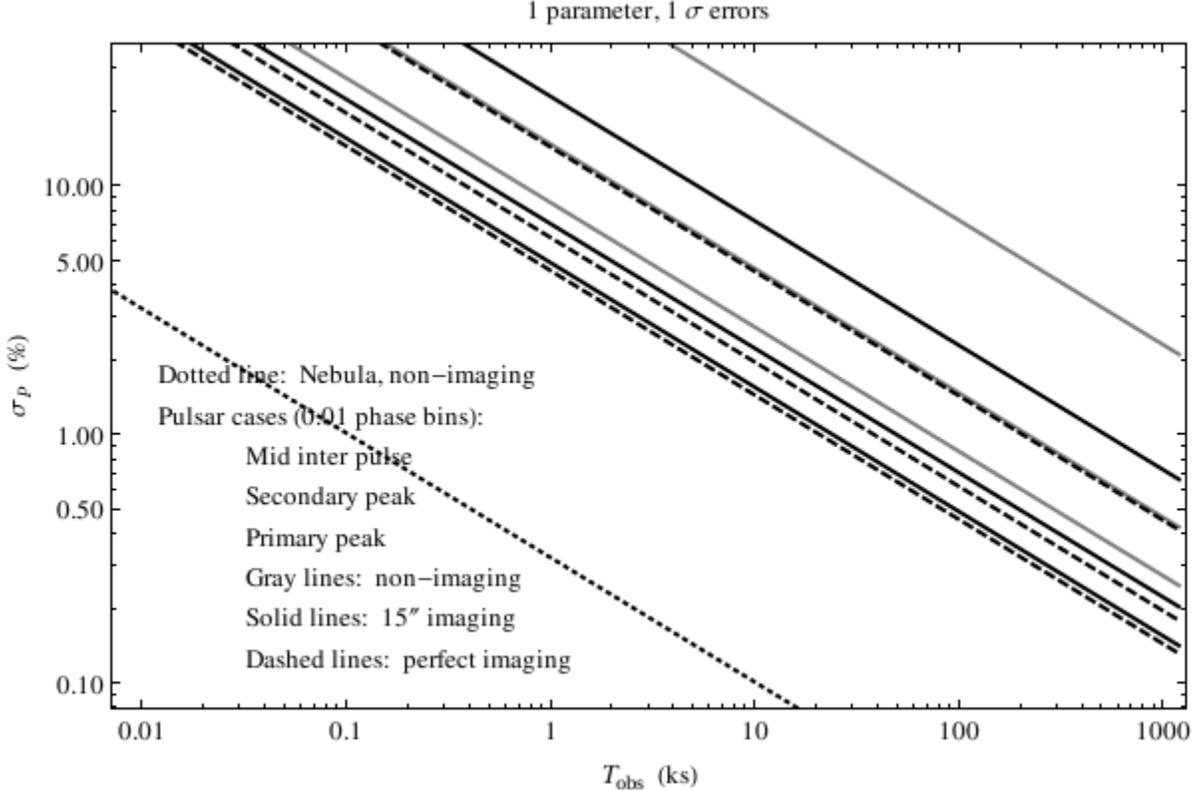

Figure 4: Univariate 1-sigma uncertainty $\sigma_{\varpi P}$ in measurement of the x-ray polarization amplitude $\varpi_P$ at representative pulse phases $\phi$ of the Crab Pulsar. As the x-ray polarization amplitude and position angle are likely to change rapidly with pulse phase (cf. Figure 2 bottom two panels), counts are accumulated in short pulse-phase bins $\phi = 0.01$. In order of decreasing polarization uncertainty $\sigma_\varpi$ (increasing sensitivity), each family of curves shows the x-ray $\sigma_{\varpi P}$ near mid inter pulse ($\phi_{IP} = 0.15$), secondary-pulse peak ($\phi_{SP} = 0.40$), and primary-pulse peak ($\phi_{PP} = 0.00$). The solid-grey, solid-black, and dashed-black lines denote non-imaging, 15-arcsec-imaging, and perfect-imaging polarimeters, respectively.

## 4. SUMMARY

We have outlined the derivation of the uncertainties relevant to x-ray polarization measurements. In so doing, we have clarified the relationship between figures of merit—the minimum detectible polarization (MDP) and the statistical accuracy of a polarization measurement. We emphasize that the integration time required to achieve a certain statistical accuracy in a polarization measurement is different from, and may be substantially longer, than the time to achieve a particular MDP. We have also examined the impact of certain systematic effects—namely, a (polarized or unpolarized) background.

Even if perfectly known through on-ground and on-orbit calibration, systematic effects may have a profound impact on sensitivity, possibly rendering certain measurements impossible. Consequently, we expect to extend our investigation to include other and more complicated systematic effects, as well as subtle corrections—such as nonuniform statistical variance—to the traditional statistical analysis.

## ACKNOWLEDGMENTS

We acknowledge and recognize Professor Robert Novick, who pioneered the field of astronomical x-ray polarimetry. For this and the more detailed study, we performed and verified symbolic calculations and manipulations using *Mathematica*®.

## APPENDIX A

A polarimeter measures linear polarization by determining the amplitude and phase of modulation with azimuthal angle $\varphi$ by analyzing the data for a $2\varphi$ signature. We write the azimuthal density of counts[a] (counts per unit azimuth) due to the source:

$$\frac{dC_S(\varphi)}{d\varphi} = \frac{C_S}{2\pi}\left[1 + a_S \cos(2(\varphi - \psi_S))\right], \tag{16}$$

with $C_S$ the total number of source counts, $a_S = \mu_S \varpi_S$ the fractional modulation amplitude for a polarization modulation factor $\mu_S$ and fractional (linear) polarization $\varpi_S$ of the source, and $\psi_S$ the corresponding polarization position angle. We write the analogous expression for the azimuthal density of counts due to the background, which may also be polarized:

$$\frac{dC_B(\varphi)}{d\varphi} = \frac{C_B}{2\pi}\left[1 + a_B \cos(2(\varphi - \psi_B))\right], \tag{17}$$

with $C_B$ the total number of background counts $a_B = \mu_B \varpi_B$ the fractional modulation amplitude for a polarization modulation factor $\mu_B$ and fractional (linear) polarization $\varpi_B$ of the background, and $\psi_B$ the corresponding polarization position angle.

Under the hypothesis that true azimuthal density of counts is

$$\frac{d\hat{C}(\varphi)}{d\varphi} = \frac{C_S}{2\pi}\left[1 + \hat{a}_S \cos(2(\varphi - \hat{\psi}_S))\right] + \frac{C_B}{2\pi}\left[1 + \hat{a}_B \cos(2(\varphi - \hat{\psi}_B))\right], \tag{18}$$

one can, in principle, derive the probability density function (PDF)

$$p(a_S, \psi_S, a_B, \psi_B \mid \hat{a}_S, \hat{\psi}_S, \hat{a}_B, \hat{\psi}_B) \tag{19}$$

for the polarization modulation parameters $a_S, \psi_S, a_B,$ and $\psi_B$, given their "true" expectation values (denoted by the caret). The derivation begins by considering $\Delta\chi^2$, the change in chi-square in varying the parameters about their expectation values, which has an azimuthal density

$$\frac{d}{d\varphi}\left[\Delta\chi^2(\varphi)\right] = \left\{\frac{dC(\varphi)}{d\varphi} - \frac{d\hat{C}(\varphi)}{d\varphi}\right\}^2 \div \frac{d\hat{C}(\varphi)}{d\varphi}. \tag{20}$$

Thus, using the expression for the azimuthal density of counts (Eq. 18), we calculate the azimuthal density of $\Delta\chi^2$:

---

[a] In sketching the derivation, we use a compact differential notation. If one bins the data into $J$ azimuthal bins, each spanning an azimuth interval $\Delta\varphi_j = (2\pi/J)$, such that the average counts per bin is $<C_j> = (C/J)$, the value of $\Delta\chi^2$ is independent of $J$. Thus we somewhat heuristically take $J \to \infty$.

$$\frac{d}{d\varphi}\left[\Delta\chi^2(\varphi)\right] = \left\{\left(\frac{C_S}{2\pi}[1+a_S\cos(2(\varphi-\psi_S))]+\frac{C_B}{2\pi}[1+a_B\cos(2(\varphi-\psi_B))]\right)\right.$$
$$\left.-\left(\frac{C_S}{2\pi}[1+\hat{a}_S\cos(2(\varphi-\hat{\psi}_S))]+\frac{C_B}{2\pi}[1+\hat{a}_B\cos(2(\varphi-\hat{\psi}_B))]\right)\right\}^2 \quad (21)$$
$$\div\left(\frac{C_S}{2\pi}[1+\hat{a}_S\cos(2(\varphi-\hat{\psi}_S))]+\frac{C_B}{2\pi}[1+\hat{a}_B\cos(2(\varphi-\hat{\psi}_B))]\right),$$

which simplifies to

$$\frac{d}{d\varphi}\left[\Delta\chi^2(\varphi)\right] = \{C_S[a_S\cos(2(\varphi-\psi_S))-\hat{a}_S\cos(2(\varphi-\hat{\psi}_S))]+C_B[a_B\cos(2(\varphi-\psi_B))-\hat{a}_B\cos(2(\varphi-\hat{\psi}_B))]\}^2$$
$$\div\{2\pi(C_S[1+\hat{a}_S\cos(2(\varphi-\hat{\psi}_S))]+C_B[1+\hat{a}_B\cos(2(\varphi-\hat{\psi}_B))])\}$$
$$= \{[C_Sa_S\cos(2(\varphi-\psi_S))+C_Ba_B\cos(2(\varphi-\psi_B))] \quad (22)$$
$$-[C_S\hat{a}_S\cos(2(\varphi-\hat{\psi}_S))+C_B\hat{a}_B\cos(2(\varphi-\hat{\psi}_B))]\}^2$$
$$\div\{2\pi([C_S+C_B]+[C_S\hat{a}_S\cos(2(\varphi-\hat{\psi}_S))+C_B\hat{a}_B\cos(2(\varphi-\hat{\psi}_B))])\}.$$

In general, the denominator, which describes the statistical variance (power in counting noise), varies with azimuthal angle $\varphi$. It is strictly constant with respect to $\varphi$ only for the double null hypothesis— $\hat{a}_S = 0$ and $\hat{a}_B = 0$ —that neither the source nor the background is polarized. However, the error in $\Delta\chi^2$ resulting from neglecting the azimuthal variation in the statistical variance is of relative order $\hat{a}^2$, which is typically <<1. Consequently, we approximate[b] the above equation:

$$\frac{d\Delta\chi^2(\varphi)}{d\varphi} \approx \frac{\{C_S[a_S\cos(2(\varphi-\psi_S))-\hat{a}_S\cos(2(\varphi-\hat{\psi}_S))]+C_B[a_B\cos(2(\varphi-\psi_B))-\hat{a}_B\cos(2(\varphi-\hat{\psi}_B))]\}^2}{2\pi(C_S+C_B)} \quad (23)$$

Here, we assume that the background is perfectly known — $a_B \to \hat{a}_B$ and $\psi_B \to \hat{\psi}_B$ — such that the remaining parameters of interest are $a_S$ and $\psi_S$:

$$\frac{d\Delta\chi^2(\varphi;a_S,\psi_S\mid\hat{a}_S,\hat{\psi}_S)}{d\varphi} \approx \frac{\{C_S[a_S\cos(2(\varphi-\psi_S))-\hat{a}_S\cos(2(\varphi-\hat{\psi}_S))]\}^2}{2\pi(C_S+C_B)}$$
$$= \frac{[a_S\cos(2(\varphi-\psi_S))-\hat{a}_S\cos(2(\varphi-\hat{\psi}_S))]^2}{2\pi\,\tilde{\sigma}_{a_S}^2}, \quad (24)$$

with $\tilde{\sigma}_{a_S} \equiv \sqrt{C_S+C_B}/C_S$ setting the scale for fitting the fractional modulation of the signal in the presence of background. Note that, as the background is assumed to be perfectly known, the modulation amplitude and the position angle of the background polarization no longer appear in the expressions for $\Delta\chi^2$ and for the consequent PDF. However, background counts still contribute to the statistical noise in measuring the modulation amplitude and, hence, will diminish the sensitivity for measuring the polarization of the source.

---

[b] Elsewhere we shall investigate consequences of the azimuthal variation of the statistical variance. As the mathematical expressions are quite complicated, we here restrict the analyses to cases for which the fractional modulation amplitude $a$ is sufficiently small that the statistical variance may be regarded as uniform.

It is straightforward to integrate the above approximation over azimuth to obtain

$$\Delta\chi^2(a_S,\psi_S \mid \hat{a}_S,\hat{\psi}_S) \approx \frac{\frac{1}{2}(a_S^2+\hat{a}_S^2)-a_S\hat{a}_S\cos(2(\psi_S-\hat{\psi}_S))}{\tilde{\sigma}_{a_S}^2} = \frac{(a_S^2+\hat{a}_S^2)-2a_S\hat{a}_S\cos(2(\psi_S-\hat{\psi}_S))}{\sigma_{a_S}^2}. \qquad (25)$$

Here,

$$\sigma_{a_S} = \sqrt{2}\,\tilde{\sigma}_{a_S} \equiv \sqrt{2(C_S+C_B)}/C_S \qquad (26)$$

sets the scale for estimating the uncertainty in $a_S$, the fractional modulation amplitude of the signal in the presence of background. This potentially confusing factor-of-√2 difference between the uncertainty in determining the modulation amplitude $\sigma_{a_S}$ and the statistical standard deviation $\tilde{\sigma}_{a_S}$ in fitting the modulation signal basically represents the difference between the amplitude and RMS (root mean square) of a sinusoidal function—$\langle \cos^2(n\theta) \rangle_{2\pi} = 1/2$.

Using the trigonometric identity $\cos(2\theta) = 1 - 2\sin^2(\theta)$, we alternatively write

$$\Delta\chi^2(a_S,\psi_S \mid \hat{a}_S,\hat{\psi}_S) \approx \frac{\frac{1}{2}(a_S-\hat{a}_S)^2 + 2a_S\hat{a}_S\sin^2(\psi_S-\hat{\psi}_S)}{\tilde{\sigma}_{a_S}^2} = \frac{(a_S-\hat{a}_S)^2 + a_S\hat{a}_S[2\sin(\psi_S-\hat{\psi}_S)]^2}{\sigma_{a_S}^2}. \qquad (27)$$

Given the approximate expression for $\Delta\chi^2$, we can now write the bivariate probability density function (PDF) $p$ in the approximation discussed above, giving the familiar[10] result:

$$\begin{aligned}
p_{a_S,\psi_S}(a_S,\psi_S \mid \hat{a}_S,\hat{\psi}_S) &\approx \frac{a_S}{\pi\sigma_{a_S}^2} \exp\left[-\tfrac{1}{2}\Delta\chi^2(a_S,\psi_S \mid \hat{a}_S,\hat{\psi}_S)\right] \\
&\approx \frac{a_S}{\pi\sigma_{a_S}^2} \exp\left[-\frac{(a_S^2+\hat{a}_S^2)-2a_S\hat{a}_S\cos(2(\psi_S-\hat{\psi}_S))}{2\sigma_{a_S}^2}\right] \\
&= \frac{a_S}{\pi\sigma_{a_S}^2} \exp\left[-\frac{(a_S-\hat{a}_S)^2 + a_S\hat{a}_S[2\sin(\psi_S-\hat{\psi}_S)]^2}{2\sigma_{a_S}^2}\right],
\end{aligned} \qquad (28)$$

where $p_{a_S,\psi_S}(a_S,\psi_S \mid \hat{a}_S,\hat{\psi}_S)$ is normalized over the intervals $0 \le a_S < \infty$ and $0 \le \psi_S < \pi$.

Integrating the bivariate PDF over the position angle $\psi_S$, we obtain the well-known univariate PDF for the modulation amplitude without regard to the position angle, described by the Rice distribution[11]:

$$\begin{aligned}
p_{a_S}(a_S \mid \hat{a}_S) &\approx \frac{a_S}{\sigma_{a_S}^2} \exp\left[-\frac{(a_S^2+\hat{a}_S^2)}{2\sigma_{a_S}^2}\right] I_0\left[\frac{a_S\hat{a}_S}{\sigma_{a_S}^2}\right] \\
&= \frac{a_S}{[2(C_S+C_B)/C_S^2]} \exp\left[-\frac{(a_S^2+\hat{a}_S^2)}{2[2(C_S+C_B)/C_S^2]}\right] I_0\left[\frac{a_S\hat{a}_S}{[2(C_S+C_B)/C_S^2]}\right],
\end{aligned} \qquad (29)$$

where $I_0(x)$ is the modified Bessel function of the first kind and order 0.

If the source is unpolarized, $\hat{a}_S \to 0$ and the PDF for obtaining a modulation amplitude $a_S$ is described by the Rayleigh distribution:

$$p_{a_S}(a_S | 0) \approx \frac{a_S}{\sigma_{a_S}^2} \exp\left[-\frac{a_S^2}{2\sigma_{a_S}^2}\right] = -\frac{d}{da_S}\left\{\exp\left[-\left(\frac{a_S^2}{2\sigma_{a_S}^2}\right)\right]\right\}. \tag{30}$$

Integrating this PDF from 0 to $\text{MDA}_{\text{CL}}$ and setting the integrand to CL, one solves for the minimum detectible (fractional) amplitude $\text{MDA}_{\text{CL}}$ at confidence level CL:

$$\text{MDA}_{\text{CL}} \equiv \sigma_{a_S}\sqrt{-2\ln(1-\text{CL})} = \frac{\sqrt{2(C_S + C_B)}}{C_S}\sqrt{-2\ln(1-\text{CL})}. \tag{31}$$

In general, integration of the PDF (Equation 28) allows exploration of the uncertainty in the parameters. If detection is significant at a high confidence level, $a_S$ and $\psi_S$ is each approximately normally distributed about its true value, such that the univariate uncertainty in the parameters for a given $\Delta\chi^2$ follows simply from Equation 27:

$$\delta a_S(\Delta\chi^2) \approx \sqrt{\Delta\chi^2} \times \sigma_{a_S} = \sqrt{\Delta\chi^2} \times \frac{\sqrt{2(C_S + C_B)}}{C_S}; \tag{32}$$

$$\delta\psi_S(\Delta\chi^2) \approx \sqrt{\Delta\chi^2} \times \frac{\sigma_{a_S}}{2\sqrt{a_S \hat{a}_S}} = \sqrt{\Delta\chi^2} \times \frac{\sigma_{a_S}}{2a_S} = \sqrt{\Delta\chi^2} \times \frac{\sqrt{2(C_S + C_B)}/C_S}{2a_S}, \tag{33}$$

where we approximate $\sin(\delta\psi) \to \delta\psi$, valid for $(\delta\psi)^2 \ll 1$. The "one-sigma" (univariate) uncertainties in the modulation parameters are then

$$\sigma_{a_S} \equiv \delta a_S(1) = \sigma_{a_S} = \frac{\sqrt{2(C_S + C_B)}}{C_S}; \tag{34}$$

$$\sigma_{\psi_S} \equiv \delta\psi_S(1) = \frac{\sigma_{a_S}}{2\sqrt{a_S \hat{a}_S}} \approx \frac{\sigma_{a_S}}{2a_S} = \frac{\sqrt{2(C_S + C_B)}/C_S}{2a_S}. \tag{35}$$

If the detection is not highly significant, the confidence contours are more complicated[12]. However, when that is the case, it is not appropriate to expect an accurate measurement.